# Leveraging AI for Enhanced Software Effort Estimation: A Comprehensive Study and Framework Proposal

Tue Nhi Tran
SGPT Department
Savvycom Software
Hanoi, Vietnam
nhi.tran@savvycomsoftware.com

Huyen Tan Tran
SGPT Department
Savvycom Software
Hanoi, Vietnam
tan.tran@savvycom.vn

Quy Nam Nguyen
SGPT Department
Savvycom Software
Hanoi, Vietnam
nam.nguyen@savvycom.vn

**Abstract**—This paper presents an extensive study on the application of AI techniques for software effort estimation in the past five years from 2017 to 2023. By overcoming the limitations of traditional methods, the study aims to improve accuracy and reliability. Through performance evaluation and comparison with diverse Machine Learning models, including Artificial Neural Network (ANN), Support Vector Machine (SVM), Linear Regression, Random Forest and other techniques, the most effective method is identified. The proposed AI-based framework holds the potential to enhance project planning and resource allocation, contributing to the research area of software project effort estimation.

## I. INTRODUCTION

In software development, accurate effort estimation is crucial for effective project planning and resource allocation. It enables project managers to make informed decisions, estimate project timelines, and allocate resources efficiently. However, traditional estimation methods often rely on expert judgment, historical data, or simplistic models, leading to suboptimal results and cost overruns.

To address these challenges, researchers and practitioners have turned to Artificial Intelligence (AI) techniques, specifically Machine Learning, to improve software effort estimation. Machine Learning algorithms have demonstrated remarkable capabilities in extracting patterns and relationships from large datasets, enabling more accurate predictions based on the inherent complexity of software projects.

This paper presents an in-depth study on the application of AI-based effort estimation techniques for software projects. Our research aims to leverage the power of Machine Learning to overcome the limitations of traditional approaches and improve the accuracy and reliability of effort estimation.

The following are the aims of this paper:

1. To review and discuss how AI techniques are used for estimating software effort.
2. To present existing challenges and principal results of software effort estimation techniques.
3. To propose an AI based Effort Estimation Framework for Software Projects.

The structure of the work is as follows. Section two provides a comprehensive literature review, section three discusses methodology of conducting the research, section four synthesizes major findings, section five illustrates the proposed framework and section six concludes.

## II. RELATED WORK

To address the existing gaps in the literature, a systematic review was implemented to examine and analyze diverse AI-based techniques utilized for estimating software effort, which include Decision Tree, Support Vector Machine (SVM), Artificial Neural Network (ANN), Linear Regression, Naive Bayes, and others.

### A. Artificial Neural Network (ANN)

The ANN model is one of the commonly used architectures in Software Cost Estimation (SCE). Researchers have proposed different ANN models to implement SCE, including Back Propagation and Evolutionary algorithms. But based on the number of research papers related to each learning algorithm, the number of studies on Back Propagation is superior because Back Propagation is used more in practice than Evolutionary algorithms. In general, the performance of ANN architecture is always better from about 91.24% to 94% when using MAE scale and a proximately 86.82% to 92.88% when using MMRE scale compared to Linear Regression (LR). It's also improves from about 56.29% to 88.1% when using MAE scale and about 26.94% to 85.33% when using the MMRE scale compared to Support Vector Machine (SVM) in each dataset used.

| Reference | Learning Technique | Learning Algorithm | Dataset | No. of Samples | No. of Variables | Performance Results |
|---|---|---|---|---|---|---|
| Ali, S. S., Ren, J., Zhang, K., Wu, J., & Liu, C. (2023) [1] | ANN and SVM and Judgement | Back Propagation | ISBSG | 951 | 11 | MMRE (0.06), MAE (302.78), MdMRE (0.0298), MdAE (21.72), PRED (0.948) |
| Rankovic, N., Rankovic, D., Ivanovic, M., & Lazic, L. (2021) [2] | Artificial Neural Network (ANN) | Back Propagation | COCOMO2000/ NASA60/ COCOMO81/ Kenner | 276 | 22 | MRE (0.4310) |
| Ali, S. S., Ren, J., Zhang, K., Wu, J., & Liu, C. (2023) [1] | Multi-layered perceptron (MLP) | Back Propagation | ISBSG | 951 | 11 | MMRE (0.022), MAE (36.23), MdMRE (0.002), MdAE (3.522), PRED (1.0) |
| Rankovic, N., Rankovic, D., Ivanovic, M., & Lazic, L. (2021) [2] | ANN18 | Back Propagation | COCOMO81/ COCOMO2000/ NASA project dataset/ Kemerer dataset | 276 | 15 | Small cluster: MAE (35.3), MMRE (14.5) Medium cluster: MAE (172.4), MMRE (162.5) Large cluster: MAE (3019.3), MMRE (1104.1) |
| | ANN27 | | | | | Small cluster: MAE (48.9), MMRE (41.9) Medium cluster: MAE (43), MMRE (38.9) Large cluster: MAE (55.6), MMRE (29.9) |
| Nagarajappa, R. K. B. & Suresh, Y. (2023) [3] | ANN | Evolutionary algorithm | COCOMO | 7399 | 16 | MSE (480.82), RMSE (21.92), MAE (10.53), MMRE (38.07) |
| Şengüneş, B., & Öztürk, N. (2023) [4] | ANN | Back Propagation | COCOMO/NASA/ Kemerer | 101 | 12 | MMRE (0.3), PRED (0.73), MAE (35.89), RMSE (45.03) |

Table 1: Overview Artificial Neural Network

| Reference | Learning Technique | Learning Algorithm | Dataset | No. of Samples | No. of Variables | Performance Results |
|---|---|---|---|---|---|---|
| Cabral, J. T. H. D. A., & Oliveira, A. L. (2021) [5] | SVR | SVR | ISBSG, PROMISE, NASA | 1468 | 20 | ISBSG: MAR (2341), PROMISE: MAR (684) |
| | BAGGING SVR | | | | | ISBSG: MAR (2589), PROMISE: MAR (709) |
| | BOOSTING SVR | | | | | ISBSG: MAR (2329), PROMISE: MAR (1003) |
| Ali, S. S., Ren, J., Zhang, K., Wu, J., & Liu, C. (2023) [1] | Support Vector Regression (SVR) | SVR | ISBSG | 951 | 11 | MMRE (0.15), MAE (304.46), MdMRE (0.099), MdAE (275.37), PRED (0.875) |
| Nagarajappa, R. K. B. & Suresh, Y. (2023) [3] | Support Vector Regression (SVR) | SVR | COCOMO | 7399 | 16 | MSE (935.84), RMSE (30.59), MAE (24.09), MMRE (52.11) |
| De Carvalho, H. D. P., Fagundes, R., & Santos, W. (2021) [6] | Support Vector Regression (SVR) | SVR | DESHARNAIS | 1000 | 12 | MMRE (0.413), MAE (0.071), MSE (0.01), RMSE (0.099) |
| Gautam, S. S., & Singh, V. (2022) [7] | SVR_BASE | Sequential Minimal Optimization (SMO) | ISBSG Release 2021/UCP/ NASA93/ China | 1179 | 10 | MAE (0.076), MSE (0.006), MdME (0.082) |
| | SVR_GSAD | | | | | MAE (0.063), MSE (0.006), MdME (0.082) |
| | SVR_IQR | | | | | MAE (0.018), MSE (0.000), MdME (0.020) |
| | RSV_Cooks | | | | | MAE (0.079), MSE (0.006), MdME (0.086) |

Table 2: Overview of Support Vector Machine

## B. Support Vector Machine (SVM)

Evgeniou et al proposed a powerful learning algorithm based on recent advances in statistical learning theory proposed. It uses a hypothesis space of linear functions in a high dimensional space, trained with a learning algorithm from optimization theory, and it implements a learning bias derived from statistical learning theory. In addition, for predicting software costs, SVR applies a linear model to implement non-linear class borders. It maps nonlinear input vectors (consisting of EM and Size of the projects) into a high dimensional attributes space by means of kernels. Overall, the proposed GSAD approach is efficient and competitive in enabling a simple yet effective outlier identification and removal procedure to improve the performance of investigated SDEE methods [7]. However, when compared with a model like ANN, there is a loss when the values of the scales like MAE and MMRE are both lost. Performance shows a decline from 56.29% to 88.1% with MAE, and from 26.94% to 85.33% with MMRE, compared to ANN.

## C. Linear Regression

As one of the most basic models in classification algorithms, the performance results returned by Linear Regression (LR) are, however, usually poor (performance is worse 91.24% when compared with MAE and 86.82% when compared with MMRE) to very poor (performance is worse 94% when compared with MAE and 92.88% when compared with MMRE) in comparison with ANN in papers that use LR for research. Linear Regression (LR) or variants such as Multiple Linear Regression (MLR) are only used in research to compare performance with other models to provide a more general overview.

| Reference | Learning Technique | Learning Algorithm | Dataset | No. of Samples | No. of Variables | Performance Results |
|---|---|---|---|---|---|---|
| Ali, S. S., Ren, J., Zhang, K., Wu, J., & Liu, C. (2023) [1] | Regression analysis | Multiple Linear Regression (MLR) | ISBSG | 951 | 11 | MMRE (0.309), MAE (603.54), MdMRE (0.2704), MdAE (342.62), PRED (0.5) |
| Nagarajappa, R. K. B. & Suresh, Y. (2023) [3] | Regression analysis | Multiple Linear Regression (MLR) | COCOMO | 7399 | 16 | MSE (17930.6), RMSE (133.9), MAE (120.18), MMRE (288.79) |
| De Carvalho, H. D. P., Fagundes, R., & Santos, W. (2021) [6] | Regression analysis | Multiple Linear Regression (MLR) | DESHARNAIS | 1000 | 12 | MMRE (0.22), MAE (0.06), MSE (0.0088), RMSE (0.092) |

Table 3: Overview of Linear Regression

| Reference | Learning Technique | Learning Algorithm | Dataset | No. of Samples | No. of Variables | Performance Results |
|---|---|---|---|---|---|---|
| Gautam, S. S., & Singh, V. (2022) [7] | RF_BASE | Decision Tree | ISBSG Release 2021/ UCP/NASA93/China | 1179 | 10 | MAE (0.018), MSE (0.001), MdME (0.009) |
| | RF_GSAD | | | | | MAE (0.013), MSE (0.000), MdME (0.007) |
| | RF_IQR | | | | | MAE (0.007), MSE (0.000), MdME (0.005) |
| | RF_Cooks | | | | | MAE (0.015), MSE (0.000), MdME (0.008) |
| Mustapha, H., & Abdelwahed, N. (2019) [8] | Random Forest based Model | Decision Tree | ISBSG/COCOMO/ TUKUTUKU | 456 | 13 | ISBSG: MMRE (1.29), MdMRE (0.37), Pred (40) |
| | | | | | | COCOMO: MMRE (1.40), MdMRE (0.55), Pred (30.43) |
| | | | | | | TUKUTUKU: MMRE (1.50), MdMRE (0.67), Pred (25) |
| Rodríguez Sánchez, E., Vázquez Santacruz, E. F., & Cervantes Maceda, H. (2023) [9] | Ensemble Learning | Random Forest | Albrecht/China/ Desharnais/Kemerer/ Kitchenham/ Maxwell/COCOMO | 21 | 12 | Time: R2 (0.9869), MRE (0.0308), RMSE (2.92), Explained Variance (0.9869) |
| | | | | | | Cost: R2 (0.9885), MRE (0.0297), RMSE (73.781), Explained Variance (0.9890) |

Table 4: Overview of Random Forest

## D. Random Forest

Approaches to problems and solving problems such as accuracy or overfitting are also mentioned in detail, for example here is the paper of the authors Rodríguez Sánchez, E., Vázquez Santacruz, E. F., & Cervantes Maceda, H. (2023). They used the Synchronous Learning technique to minimize bias by combining individual models and their predictions. Although a set that includes the base model requires more computation, it also reduces method error because the mean of the set will come close to the target value. Overall, the results of Random Forest are good when comparing R2, MMRE and Pred (approximately 7.19%) with single SVM [9], there are also effective methods as a Decision Tree was the base learner for the Random Forest and AdaBoost to reduce bias and increase accuracy, thereby increasing efficiency for output results.

## E. Others

Some papers mention other models along with other Learning algorithms to test their feasibility in the Software Cost Estimation (SCE) problem.

A rather interesting approach as suggested by author Tawosi is to use Deep-SE as a deep learning model for estimation. Their model consists of four components: Word Embedding, Document Representation using Long-Term Memory (LSTM), Deep Representation using Regression Highway Network (RHWN), and Differential Regression, or another approach such as using Genetic Algorithm (GA) to find the most expensive set of CBR (Cased-Based Reasoning technique) parameters.

However, the above methods only give good performance in some cases, such as when Deep-SE is used on non-transformed data (Deep-SE achieves lower MAE values than both baseline techniques for all 16 projects from 2.86% to 35.67%) or when the CBR method is optimized by GA in large datasets that are mostly numeric and have sufficient projects and features to allow learning (CBR-GA was better than other models with large number of wins that is 85.9%). In other cases, such as crossing-project estimation (worse 4% with baseline techniques) or analyzing small datasets that have a few numbers of observations and low dimensionality in CBR-GA (CBR-GA was better than other models with small number of wins that is 40%), they give non-improving results. Therefore, it is necessary to consider carefully before using them.

| Reference | Approach | Learning Algorithm | Dataset | No. of Samples | No. of Variables | Performance Results |
|---|---|---|---|---|---|---|
| Hameed, S., Elsheikh, Y., & Azzeh, M. (2023) [10] | CBR-GA model | Genetic Algorithm | Albrecht/China/ Desharnais/ Kemerer/ Kitchenham/ Maxwell/COCOMO | 713 | 18 | MAE (7.742), MBRE (80.94), MIBRE (32.952) |
| Tawosi, V., Moussa, R., & Sarro, F. (2022) [11] | Deep-SE and the baseline estimators | Word Embedding, Document representation using Long-Short Term Memory (LSTM), Deep representation using Recurrent Highway Network (RHWN) and Differentiable Regression. | Choetkiertikul/ Porru/Tawosi | 76 | 14 | MAE (1.36) |
|  | Deep-SE and TF/IDF-SVM |  |  |  |  | MAE (1.04) |
|  | Deep-SE with Cross-project Estimation |  |  |  |  | MAE (1.99) |
|  | Augmented Training Set with Deep-SE |  |  |  |  | MAE (4.31) |
|  | Deep-SE with Pre-Training Effectiveness |  |  |  |  | MAE (1.212) |

Table 5: Overview of others models

## III. METHODOLOGY

### 1. Literature meta-analysis method

Our research team searched for relevant articles related to research topics, and we applied criteria to filter and select appropriate articles for analysis.

The processes of literature meta-analysis are as follows:

- The five academic databases used to collect articles are IEEE Xplore, ACM, Springer, Scopus and Science Direct.
- The keywords used for literature search are "software estimation efforts" OR "estimate software efforts" AND ("AI" OR "Artificial Intelligence" OR "Machine Learning")
- The time range of articles published from January 2017 to December 2023 will allow us to explore the most recent techniques in the past five years.
- After selecting academic papers according to the criteria, we finally obtained the 22 most relevant research papers.

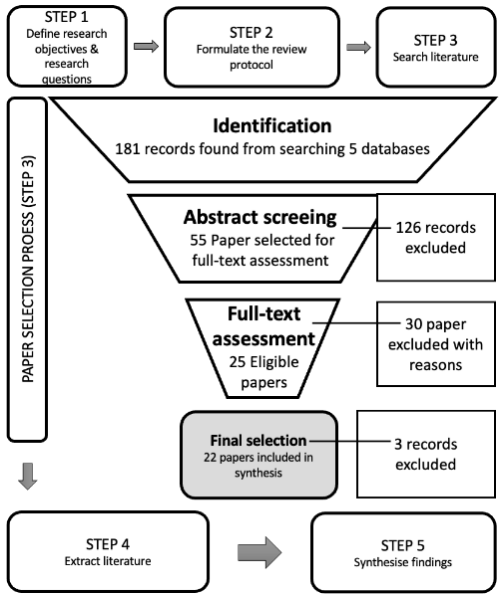

Figure 1: Systematic review methodology

### 2. ANN method

ANN is used efficiently in these studies because it can learn and synthesize information from the input data features and build a valuable predictive model. Also, ANN uses Back Propagation for optimizing, which brings positive results. Based on these reasons, it can be said that ANN with Back Propagation is currently the best model to use for solving estimation problems.

We chose the combination of Backpropagation and MLP to build the estimation model as we can gain many benefits such as increasing the accuracy of the model, reducing the training time, dealing with nonlinear features, and easy deployment. MLP can learn complex features of data, and Backpropagation helps to optimize the weights of neurons in the network. Combining these two technologies makes training models faster and more efficient, handling nonlinear features and easy deployment.

## IV. PRELIMINARY RESULTS

### 1. Overview of Machine Learning techniques for estimating software efforts

In this paper, we present a review of different Machine Learning techniques used in software effort estimation. We chose 22 numbers of papers that have been used for software effort estimation by applying different techniques to improve the accuracy of estimated output. Efforts are made to critically examine the literature based on several criteria, including sample size, number of variables utilized in the study, validation methods employed, and the measure used for comparing the performance of different techniques. The techniques discussed in this paper are summarized in Table 1 to Table 5. Each table gives a summary of different techniques at the order of Artificial Neural Network, Support Vector Machine, Linear Regression, Random Forest, and other Machine Learning techniques. As the review points out the usage of a variety of validation methods and error measures for comparison purpose, this information is also presented each study along with other relevant details in the summary tables.

Out of 22 studies, Artificial Neural Network (ANN) techniques appears in 6 papers, Support Vector Machine (SVM) in 9 researchs, Linear Regression (LR) in 5 papers, Random Forest (RF) in 6 researchs and other techniques in 8 papers. The results of comparison are presented in Figure 2.

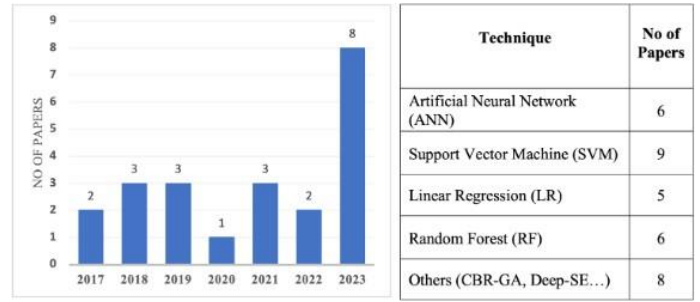

Figure 2: 22 papers selected according to time and techniques

### 2. Major inputs

The Figure 3 presents a summary of commonly used input in the dataset for the software project.

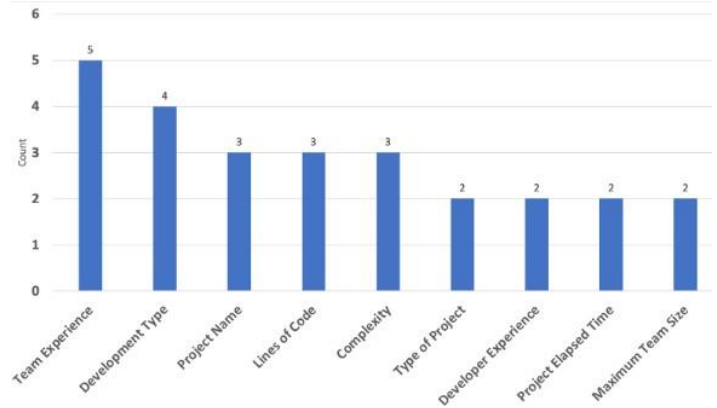

Figure 3: Major input for datasets in estimating software efforts

### 3. Major performance measures

Figure 4 presents a summary of evaluation measures used in the literature to compare the performance of different Machine Learning techniques.

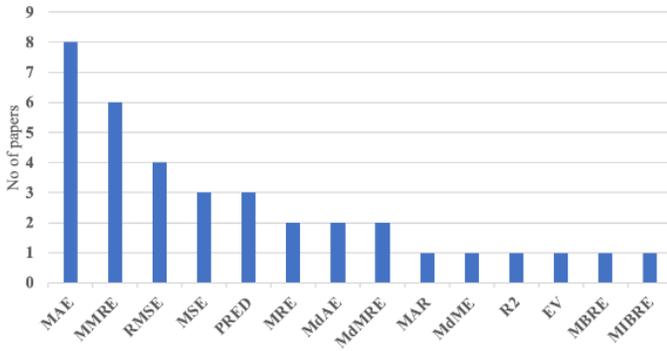

Figure 4: Major performance measures for software estimation

## V. PROPOSED WORK

The work proposes to build a framework for Software Effort Estimation (SEE) using ANN with Back Propagation, in order to optimize the estimation process and improve prediction accuracy. Data about the effort of software projects is aggregated and collected. It is a metric used to estimate the size and complexity of a software project and to measure developer productivity. These are the common inputs that are frequently used in the datasets that the research mentioned in Figure 3.

This data will be used to train the Machine Learning model. Using the training data, ANN is used to predict the expected software effort based on the collected factors. The trained Machine Learning model is implemented to estimate the software effort for new projects based on the given factors such as Team Experience, Development Type, Project Features, etc. These estimates will provide important information for quantifying the work and resources for the project total adaptive effort of project and project elapsed times. This step is shown in Figure 5.

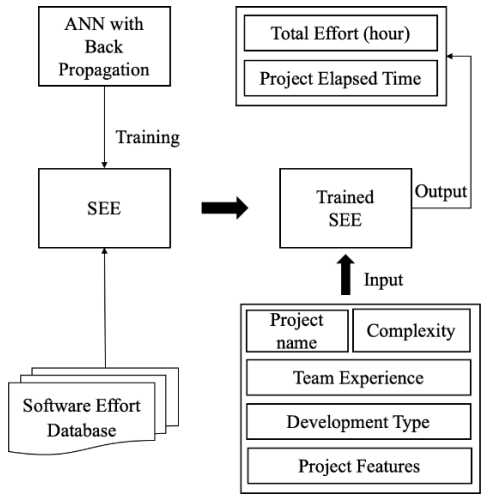

Figure 5: Training framework for software effort estimation

The proposed develop an application explanatory framework to estimate software effort using Machine Learning techniques to improve the accuracy and efficiency of the software effort estimation process and update the database. Data collection, Machine Learning model building, model evaluation and optimization will ensure the accuracy and reliability of the software effort estimate. This adaptive framework will bring benefits in quantifying the project, optimizing the process, and reducing the time, effort for software developers.

## VI. CONCLUSION

This research paper provides a comprehensive review of recent Machine Learning techniques for software effort estimation and highlights the Artificial Neural Network (ANN) as the most suitable model, given its superior performance in terms of metrics like MAE and MMRE when compared to alternative models like Support Vector Machine or Linear Regression. By implementing this framework, we anticipate significant improvements in predictability and manageability throughout the software development process, enabling better decision-making and enhanced project outcomes. As the field of Machine Learning continues to advance, the findings and methodology presented in this study can serve as a valuable reference for future research and practical applications in software engineering and project management domains.

REFERENCES

[1] Ali, S. S., Ren, J., Zhang, K., Wu, J., & Liu, C. Heterogeneous Ensemble Model to Optimize Software Effort Estimation Accuracy. IEEE Access, 11:27759-27792, 2023.

[2] Rankovic, N., Rankovic, D., Ivanovic, M., & Lazic, L. A new approach to software effort estimation using different Artificial Neural Network architectures and Taguchi orthogonal arrays. IEEE access, 9:26926-26936, 2021.

[3] Ravi Kumar B N, Yeresime Suresh. Software Development Effort Estimation Using Relational Database and Optimized Learning Mechanism. Journal of Computer Science, 19:540-553, 2023.

[4] Şengüneş, Burcu, and Nursel Öztürk. An Artificial Neural Network Model for Project Effort Estimation. Systems, 11.2:91, 2023.Cabral, Jose Thiago H. de A., and Adriano LI Oliveira. Ensemble Effort Estimation using dynamic selection. Journal of Systems and Software, 175:110904, 2021.

[5] Cabral, Jose Thiago H. de A., and Adriano LI Oliveira. Ensemble Effort Estimation using dynamic selection. Journal of Systems and Software, 175:110904, 2021.

[6] De Carvalho, Halcyon Davys Pereira, Roberta Fagundes, and Wyllians Santos. Extreme learning machine applied to software development effort estimation. IEEE Access, 9:92676-92687, 2021.

[7] Gautam, Swarnima Singh, and Vrijendra Singh. Adaptive Discretization Using Golden Section to Aid Outlier Detection for Software Development Effort Estimation. IEEE Access, 10:90369-90387, 2022.

[8] Mustapha, Hain, and Namir Abdelwahed. Investigating the use of Random Forest in software effort estimation. Procedia computer science, 148:343-352, 2019.

[9] Rodríguez Sánchez, Eduardo, Eduardo Filemón Vázquez Santacruz, and Humberto Cervantes Maceda. Effort and Cost Estimation Using Decision Tree Techniques and Story Points in Agile Software Development. Mathematics, 11.61477, 2023.

[10] Hameed, Shaima, Yousef Elsheikh, and Mohammad Azzeh. An optimized case-based software project effort estimation using genetic algorithm. Information and Software Technology, 153:107088, 2023.

[11] Tawosi, Vali, Rebecca Moussa, and Federica Sarro. Agile Effort Estimation: Have We Solved the Problem Yet? Insights From a Replication Study. IEEE Transactions on Software Engineering, 49.4:2677-2697, 2022